\documentclass[a4,10pt,twoside]{article}

\usepackage[utf8]{inputenc}
\usepackage[english]{babel}
\usepackage[T1]{fontenc}
\usepackage{url}
\usepackage[colorlinks=true, allcolors=blue]{hyperref}
\usepackage{graphicx}
\usepackage[left=85pt, right=85pt, top=85pt, bottom=95pt,
headheight=15pt,%
headsep=10pt]{geometry}%
\usepackage{amsmath, amssymb}
\usepackage{authblk}
\usepackage{doi}
\usepackage{newtxtext,newtxmath}
\usepackage{microtype}
\usepackage{booktabs}
\usepackage[bf,small]{caption}
\usepackage{tikz}
\usetikzlibrary{decorations.pathmorphing, arrows.meta}

\usepackage[figurename=Fig.]{caption}

\title{Inverse design of a spatial demultiplexer for free-space
  optical communications: direct optimization over turbulence
  statistics}

\author[1,*]{Nicolas Barré}

\affil[1]{Independent researcher}

\affil[*]{email: nicolas.barre@protonmail.com}

\date{\today}

\begin{document}

\maketitle

\begin{abstract}
Atmospheric turbulence severely limits the coupling of received optical wavefronts into
single-mode fibers in satellite-to-ground free-space optical links. Spatial demultiplexing
receivers address this challenge by distributing the incoming field across a bundle of
single-mode fibers whose outputs are recombined coherently, relaxing the requirements on
wavefront correction. In this work, we investigate the design of such receivers from two
complementary angles. We first compare the power coupling statistics achieved by several
modal bases and show that the spatial support of the modes matters far more than the
specific choice of basis, questioning the relevance of mode-selective approaches for this
application. We then present the inverse design of a compact two-plane refractive system
optimized directly over an ensemble of turbulence realizations using stochastic gradient
descent, with no constraint imposed on the input modal decomposition. The optimized system
significantly improves over direct coupling into the fiber bundle, approaches the
performance of an ideal modal projection, and remains competitive across a broad range of
turbulence conditions.
\end{abstract}

\section{Introduction}

Free-space optical (FSO) communications offer a promising alternative to radio-frequency
links for high-throughput ground-to-satellite and inter-building
connectivity~\cite{Guiomar:2022}. A central challenge in FSO systems is the coupling of the
received optical wavefront into single-mode fibers, which is severely degraded by
atmospheric turbulence. Turbulence introduces wavefront distortions that reduce the spatial
coherence of the received field, leading to coupling losses and signal fading that directly
limit the link budget.

Two main strategies have been developed to mitigate this effect. The first relies on
adaptive optics, where a wavefront sensor and a deformable mirror correct the incoming
wavefront in real time to restore single-mode fiber coupling~\cite{Paillier:2020,
  Horst:2023}. The second strategy, which is the focus of this work, relies on spatial
demultiplexing: the turbulent wavefront is distributed across a set of single-mode fibers,
and the outputs are recombined coherently to recover the transmitted
signal~\cite{Billaud:2019, Billault:21, demarinis2025, Billault:25}.

Two main device architectures have been explored to implement this spatial demultiplexing.
Photonic lanterns, originally developed for astronomical
instrumentation~\cite{norris2019astrophotonics}, interface a bundle of single-mode fibers
with the incoming multimode beam through an adiabatic waveguide taper~\cite{Birks:15}.
Ultrafast laser inscription has since extended this concept to complex three-dimensional
geometries~\cite{gross2015ultrafast}, improving robustness and design flexibility.
Multi-plane light conversion (MPLC), on the other hand, realizes a fully controlled unitary
transformation between a prescribed set of orthogonal spatial modes through successive
free-space phase modulations~\cite{Morizur:10, zhang2023multi, A.Rocha:2025}. A recent
experimental comparison over a simulated GEO-to-ground FSO link has shown that a seven-port
photonic lantern achieves on average $1\,$dB lower coupling losses than a commercial 15-mode
MPLC module, at the cost of higher signal fading~\cite{Billault:25}.

The design of multi-element diffractive optical systems, in which a wavefront undergoes
successive phase modulations separated by free-space propagation, has a long history
predating the optical neural network literature. As early as 1993, Fienup established
gradient-based inverse design of cascaded phase elements as a viable computational
approach~\cite{Fienup:93}. Wang and Piestun later applied equivalent functionality using
cascaded planar elements optimized through iterative projection algorithms, in the context
of volumetric holography and multicolor beam shaping~\cite{Wang:18}. Independently, the
optical neural network community framed the same physical architecture as an end-to-end
differentiable computing substrate, optimized for arbitrary objective
functions~\cite{Lin:2018, Mengu:2019, Zhou:2021, yildirim2024nonlinear}. Within this
framework, MPLC can be viewed as a particular instantiation that uses phase masks as optical
elements and restricts the transformation to a unitary mapping between a fixed set of
orthogonal input and output spatial modes, a natural choice for telecommunications where
mode orthogonality defines the independent channels that carry information. The approach
presented here is fundamentally different: no input modal basis is defined, and no
orthogonal mode decomposition is imposed. Instead, the optimization criterion is directly
the total power collected into the single-mode fibers, averaged over an ensemble of
atmospheric turbulence realizations, which is the physical quantity of interest in
free-space optical communications.

In this work, we approach the design of FSO spatial demultiplexers from two complementary
angles. We first systematically compare the power coupling distributions achieved by several
modal bases for turbulent wavefront collection, and show that the choice of detection basis
has a limited impact on power collection efficiency, provided the spatial support of the
modes covers the collecting pupil efficiently. We then present the inverse design of a
compact two-plane refractive system optimized directly over an ensemble of atmospheric
turbulence realizations using stochastic gradient descent, without any constraint on the
input modal decomposition. The optimized system approaches the performance of an ideal
Karhunen-Loève modal projection and significantly improves over direct coupling into the
fiber bundle, while remaining competitive across a broad range of turbulence conditions.

\section{Modal basis comparison for turbulent wavefront power coupling}
\label{sec:basis}

The fraction of power that can be collected from a turbulent wavefront depends on the choice
of modal basis onto which the incoming field is projected. Before designing any physical
system, it is therefore instructive to compare the power coupling distributions achieved by
different modal bases, independently of any hardware constraint. This section establishes
the theoretical upper bound set by the Karhunen-Lo\`eve basis, introduces the statistical
characterization methodology used throughout this work, and compares the performance of
several practically relevant modal bases.

\subsection{Turbulence statistics and Karhunen-Lo\`eve basis}

Atmospheric turbulence is modeled using a modified Von K\'arm\'an power spectrum
as defined in~\cite{mnras:2025}:
\begin{equation}
  \label{eq:von_karman}
  \Phi_\phi(u) = 0.0299 \cdot r_0^{-5/3} \cdot
  \frac{\exp\!\left(-u^2/u_o^2\right)}{\left(u^2 + u_i^2\right)^{11/6}},
\end{equation}
where $\Phi_\phi$ is the two-dimensional power spectral density of the wavefront phase, $u$
is the spatial frequency in cycles per meter, $r_0$ is the Fried parameter, and
$u_i = 1/l_0$ and $u_o = 1/L_0$ are the spatial frequencies associated with the inner scale
$l_0$ and outer scale $L_0$, respectively.

In practice, FSO receivers operate with collecting apertures ranging from a few centimeters
to tens of centimeters. Rather than simulating at the physical scale of the receiver, we
work in a reduced coordinate system where the pupil diameter is set to
$D = 160\,\mathrm{\mu}$m, corresponding to the assumption that a telescope images the
collecting aperture onto the input of the demultiplexer. All results depend only on the
dimensionless ratio $D/r_0$, which fully characterizes the turbulence strength relative to
the aperture size, and are therefore independent of the physical scale of the simulation.

Phase screens are generated on a $300 \times 300$ grid with sampling
$\Delta x = 1\,\mathrm{\mu}$m, giving a simulation domain of $300\,\mathrm{\mu}$m, slightly
larger than the pupil to avoid Fourier artifacts at the pupil boundary.  The inner scale is
set by the grid sampling, $l_0 = \Delta x = 1\,\mathrm{\mu}$m.  To extend the outer scale
and suppress low-frequency artifacts inherent to FFT-based generation, the grid is
zero-padded by a factor of 8 prior to spectral filtering, giving an effective outer scale
$L_0 = 8 \times 300\,\mathrm{\mu}$m $= 2400\,\mathrm{\mu}$m. A set of $N = 3000$ independent
turbulent phase screens is generated at $D/r_0 = 8$, corresponding to moderate turbulence
conditions representative of a ground-to-satellite FSO link.

\begin{figure}[htbp]
  \centering
  \includegraphics{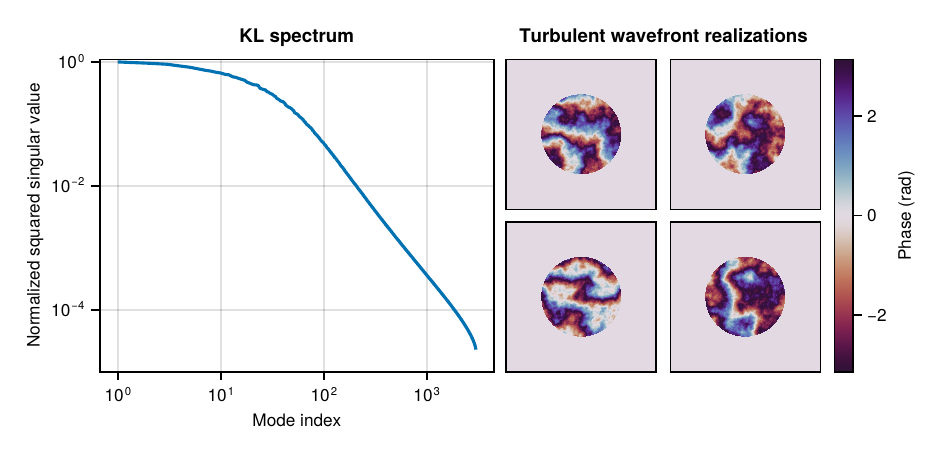}
  \caption{Left: normalized squared singular value spectrum of the
  Karhunen-Lo\`eve basis, showing the rapid decay of the eigenvalues.
  Right: four representative turbulent wavefront realizations at
  $D/r_0 = 8$, displayed within the circular collecting pupil.
  The phase is shown in radians.}
  \label{fig:turbulence}
\end{figure}

To identify the optimal modal basis for power coupling, we arrange the $N = 3000$ turbulent
wavefront realizations as columns of a data matrix $X \in \mathbb{R}^{N_{\rm pix} \times N}$
with $N_{\rm pix} = 300 \times 300$, where each realization is masked by the circular
collecting pupil prior to vectorization. The left singular vectors of $X$, obtained by
singular value decomposition, form the Karhunen-Loève (KL) basis adapted to the turbulence
statistics. Fig.~\ref{fig:turbulence} shows the normalized power spectrum of the generated
phase screens alongside representative turbulent wavefront realizations at $D/r_0 = 8$,
confirming that the simulated turbulence follows the expected Von K\'arm\'an statistics.

\begin{figure}[htbp]
  \centering
  \includegraphics{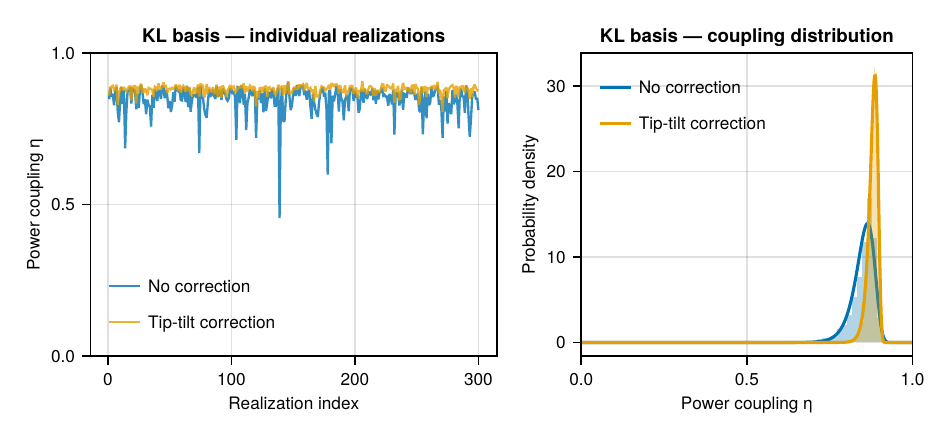}
  \caption{Power coupling efficiency $\eta$ of the Karhunen-Lo\`eve basis
  at $D/r_0 = 8$, without (blue) and with (orange) tip-tilt correction.
  Left: $\eta$ over 300 representative turbulent realizations.
  Right: distribution of $\eta$ over 3000 realizations, with fitted
  Kumaraswamy distributions overlaid.}
  \label{fig:kl_distribution}
\end{figure}

By construction, this basis maximizes the power captured by a truncated set of modes and
serves as the theoretical upper bound for any fixed modal decomposition.  We retain the
first 91 KL modes, which capture the dominant fraction of the turbulent wavefront energy for
the considered $D/r_0$ range. For each turbulent realization, we compute the total power
fraction $\eta \in [0, 1]$ collected into the 91 KL modes. Fig.~\ref{fig:kl_distribution}
shows the resulting time series of $\eta$ over 300 representative realizations, as well as
the full distribution over the 3000 realizations, both without and with tip-tilt correction
applied to the incoming wavefront prior to projection. Tip-tilt correction is considered
here as a practically relevant pre-processing step, as fast tip-tilt mirrors are routinely
deployed in FSO systems to compensate beam wander at low cost and without the complexity of
full adaptive optics correction. The distribution is well described by a Kumaraswamy
distribution, a flexible two-parameter model for bounded random variables whose asymmetric
shape, with a tail toward low values, is better suited to the observed coupling statistics
than a symmetric model such as the Beta distribution. Its probability density function reads
\begin{equation}
  f(\eta; a, b) = ab\,\eta^{a-1}(1-\eta^a)^{b-1}, \quad \eta \in [0,1],
  \label{eq:kumaraswamy}
\end{equation}
allowing the mean coupling efficiency and fading statistics to be characterized by just two
parameters. This fitting methodology is used consistently for all modal bases considered in
this work. Tip-tilt correction provides a modest improvement in mean coupling efficiency
($\bar{\eta} = 0.850$ without correction, $\bar{\eta} = 0.880$ with correction), but
significantly reduces the standard deviation of the coupling fluctuations ($\sigma = 0.033$
without, $\sigma = 0.015$ with), thereby mitigating fading events by a factor of
approximately two in standard deviation.

\subsection{Alternative modal bases and power coupling distributions}

\begin{figure}[htbp]
  \centering
  \includegraphics{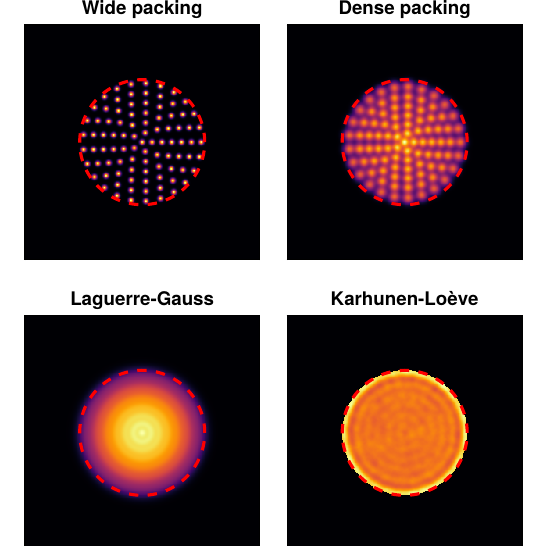}
  \caption{Total intensity of the four modal bases considered in this work, summed over the
    first 91 modes: wide packing (top left), dense packing (top right), Laguerre-Gauss
    (bottom left), and Karhunen-Lo\`eve (bottom right).  The dashed red circle indicates the
    boundary of the collecting pupil of diameter $D = 160\,\mathrm{\mu}$m.}
  \label{fig:bases_intensities}
\end{figure}

We define three additional modal bases and evaluate their power coupling distributions using
the same methodology. The first is a 91-mode Laguerre-Gaussian (LG) basis, with waist
parameter $w_{LG} = 24.5\,\mathrm{\mu}$m optimized to maximize the mean power coupling over
the turbulence ensemble. This basis is a standard reference in spatial division multiplexing
and represents a natural fixed analytical choice requiring no prior knowledge of the
turbulence statistics. The second and third bases share a common concentric ring layout:
modes are arranged in a central spot surrounded by 6 concentric rings, each ring containing
4 more equally-spaced spots than the preceding ring, starting with 5 spots on the innermost
ring, for a total of 91 modes including the central spot. In the wide packing variant, modes
have waist $w_G = 4.2\,\mathrm{\mu}$m with a center-to-center separation of $3w_G$, ensuring
approximate orthogonality of the individual modes. In the dense packing variant, the waist
is increased to $w_G^\prime = 11.5\,\mathrm{\mu}$m with a center-to-center separation of
$w_G^\prime$, resulting in a strongly non-orthogonal set that is subsequently
re-orthonormalized using a Gram-Schmidt procedure. The total intensity summed over the 91
modes of each basis is shown in Fig.~\ref{fig:bases_intensities}, confirming that all four
bases provide good spatial coverage of the collecting pupil.

\begin{figure}[htbp]
  \centering
  \includegraphics{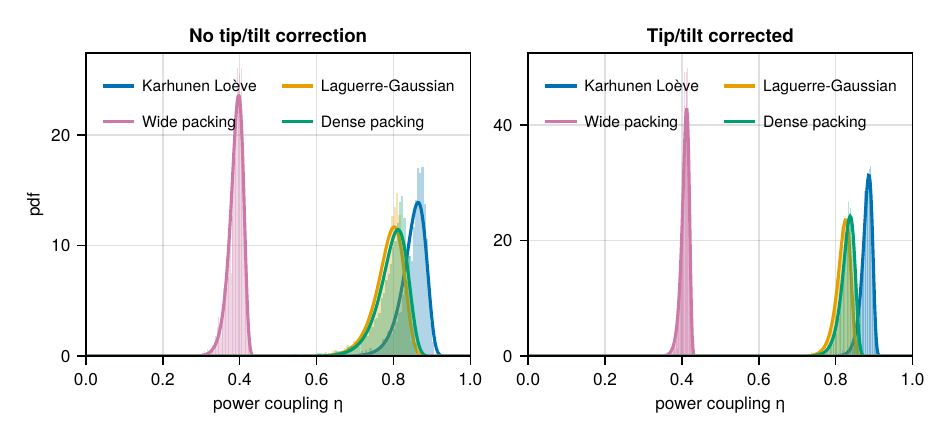}
  \caption{Power coupling efficiency distributions for the four modal bases considered in
    this work at $D/r_0 = 8$. Left: without tip-tilt correction. Right: with tip-tilt
    correction. Histograms show the empirical distributions over 3000 turbulent
    realizations; solid curves show the fitted Kumaraswamy distributions.}
  \label{fig:bases_distributions}
\end{figure}

The power coupling distributions for all four bases, fitted with Kumaraswamy distributions,
are shown in Fig.~\ref{fig:bases_distributions} and summarized in
Table~\ref{tab:basis_comparison}, for both the uncorrected and tip-tilt corrected cases.
The KL basis achieves the highest mean coupling efficiency by construction, reaching
$\bar{\eta} = 0.850$ without correction and $\bar{\eta} = 0.880$ with tip-tilt
correction. The LG basis and the dense packing basis perform comparably, with mean
efficiencies within $1.5\%$ of each other in both cases and a slight advantage for dense
packing. To further characterize the relationship between these bases, we compute their
pairwise power coupling overlap, defined as
\begin{equation}
  O(u, v) = \frac{1}{N} \sum_{i,j} |\langle u_i | v_j \rangle|^2,
  \label{eq:overlap}
\end{equation}
where $\{u_i\}$ and $\{v_j\}$ are two orthonormal bases of $N$ modes. This
quantity equals 1 when the two bases span the same subspace and measures the
fraction of power transferable between them. The resulting overlap matrix is given
in Table~\ref{tab:basis_overlap}. Both the LG and dense packing bases recover
approximately $80\%$ of the KL subspace, consistently with their similar mean
coupling efficiencies. Their mutual overlap of $86\%$ further indicates that despite
their near-equivalent performance, they do not span exactly the same functional
space and collect partly different turbulence modes. The dense packing basis,
re-orthonormalized from a set of overlapping Gaussians, spans a similar functional
space to that of fs-laser written photonic lanterns, where closely-spaced fiber
cores merge in the tapered section. In both cases the individual mode shapes are not
precisely controlled, yet the near-equivalence with LG performance confirms that
what matters is that the spatial span of the basis covers the collecting pupil
efficiently, rather than the precise shape of its individual modes.

\begin{table}[htbp]
  \centering
  \caption{Mean power coupling efficiency $\bar{\eta}$ and standard
  deviation $\sigma$ for the four modal bases, without and with tip-tilt
  correction, at $D/r_0 = 8$.}
  \label{tab:basis_comparison}
  \begin{tabular}{lcccc}
    \toprule
    & \multicolumn{2}{c}{No correction} 
    & \multicolumn{2}{c}{Tip-tilt corrected} \\
    \cmidrule(lr){2-3} \cmidrule(lr){4-5}
    Basis & $\bar{\eta}$ & $\sigma$ & $\bar{\eta}$ & $\sigma$ \\
    \midrule
    Karhunen-Lo\`eve & 0.850 & 0.033 & 0.880 & 0.015 \\
    Laguerre-Gauss & 0.786 & 0.038 & 0.818 & 0.020 \\
    Wide packing   & 0.390 & 0.019 & 0.408 & 0.011 \\
    Dense packing  & 0.796 & 0.039 & 0.830 & 0.019 \\
    \bottomrule
  \end{tabular}
\end{table}

\begin{table}[htbp]
  \centering
  \caption{Pairwise power coupling overlap $O(u, v)$ as defined in
    Eq.~\ref{eq:overlap} between the Karhunen-Loève, Laguerre-Gauss, and dense packing
    bases.}
  \label{tab:basis_overlap}
  \begin{tabular}{lccc}
    \toprule
    & KL & LG & Dense packing \\
    \midrule
    KL           & 1.000 & 0.790 & 0.806 \\
    LG           & 0.790 & 1.000 & 0.861 \\
    Dense packing & 0.806 & 0.861 & 1.000 \\
    \bottomrule
  \end{tabular}
\end{table}

The wide packing basis, where fiber modes do not overlap, represents the performance
achievable with a simple fiber bundle and a microlens array requiring no optical
optimization, and reaches only $\bar{\eta} \approx 0.39$, a factor of two below the other
bases. Interestingly, it also exhibits the lowest coupling standard deviation
($\sigma = 0.019$ without correction), as the spatial diversity of non-overlapping modes
averages out turbulence fluctuations at the cost of reduced mean efficiency. This is the
operating principle of multi-aperture receivers, here quantified in direct comparison with
modal bases of comparable complexity. Tip-tilt correction provides a modest but consistent
improvement in mean coupling efficiency across all bases ($+3$ to $+5$ percentage points),
while more significantly reducing the standard deviation of the coupling fluctuations by
approximately a factor of two, which is practically relevant for FSO link reliability.
These results motivate the inverse design approach developed in the following section, where
the output mode arrangement is no longer constrained to any predefined basis and power
coupling over turbulence statistics is maximized directly.

\section{Inverse design of a two-plane free-space to fiber bundle demultiplexer}
\label{sec:design}

The previous section established that the choice of modal basis has a limited impact on
power coupling efficiency, provided the spatial support of the modes is well matched to
the collecting pupil. This motivates a different question: given a fixed physical
architecture, can direct optimization over turbulence statistics yield a system that
approaches the theoretical upper bound set by the Karhunen-Loèeve basis, without
any reference to an input modal decomposition?

\subsection{Optical system description}

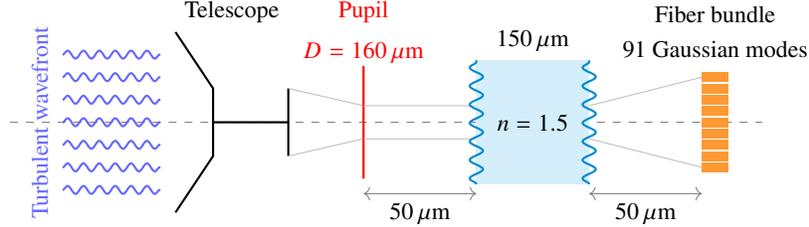
\begin{figure}[htbp]
\centering
\begin{tikzpicture}[scale=1.0, every node/.style={font=\small}]

\draw[dashed, gray] (-8.2, 0) -- (1.8, 0);

\foreach \y in {-0.9, -0.6, -0.3, 0.0, 0.3, 0.6, 0.9} {
    \draw[blue!60, thick, decorate, 
          decoration={snake, amplitude=1.5pt, segment length=6pt}]
        (-7.5, \y) -- (-6.2, \y);
}
\node[blue!60, rotate=90] at (-7.8, 0) {Turbulent wavefront};

\draw[thick] (-6, -1.2) -- (-5.5, -0.45) -- (-5.5, 0.45) -- (-6, 1.2);
\draw[thick] (-4.5, -0.45) -- (-4.5, 0.45);
\draw[thick] (-5.5, 0) -- (-4.5, 0);
\node[above] at (-5.25, 1.2) {Telescope};

\draw[gray!50] (-4.5, -0.45) -- (-3.5, -0.22);
\draw[gray!50] (-4.5,  0.45) -- (-3.5,  0.22);

\draw[thick, red] (-3.5, -0.75) -- (-3.5, 0.75);
\node[above, red] at (-3.5, 1.2) {Pupil};
\node[below, red] at (-3.5, 1.2) {$D=160\,\mathrm{\mu}$m};

\draw[gray!50] (-3.5, -0.22) -- (-2.0, -0.22);
\draw[gray!50] (-3.5,  0.22) -- (-2.0,  0.22);

\fill[cyan!15] (-2.0, -0.825) rectangle (-0.5, 0.825);
\draw[thick, cyan!70!blue, decorate, 
      decoration={snake, amplitude=2.5pt, segment length=7pt}]
    (-2.0, -0.825) -- (-2.0, 0.825);
\draw[thick, cyan!70!blue, decorate, 
      decoration={snake, amplitude=2.5pt, segment length=7pt}]
    (-0.5, -0.825) -- (-0.5, 0.825);
\node[above] at (-1.25, 0.825) {$150\,\mathrm{\mu}$m};
\node at (-1.25, 0) {$n=1.5$};

\draw[gray!50] (-0.5, -0.22) -- (1.0, -0.6);
\draw[gray!50] (-0.5,  0.22) -- (1.0,  0.6);

\foreach \y in {-0.6, -0.45, -0.3, -0.15, 0.0, 0.15, 0.3, 0.45, 0.6} {
    \fill[orange!80] (1.0, \y-0.06) rectangle (1.35, \y+0.06);
}
\node[above] at (1.175, 1.2) {Fiber bundle};
\node[below] at (1.175, 1.2) {91 Gaussian modes};

\draw[<->, gray] (-3.5, -0.95) -- (-2.0, -0.95);
\node[below] at (-2.75, -0.95) {$50\,\mathrm{\mu}$m};

\draw[<->, gray] (-0.5, -0.95) -- (1.0, -0.95);
\node[below] at (0.25, -0.95) {$50\,\mathrm{\mu}$m};

\end{tikzpicture}

\caption{Schematic of the two-plane refractive demultiplexer. A telescope images the
  collecting aperture onto a $160\,\mathrm{\mu}$m input pupil. The optimized glass slab
  ($n = 1.5$, thickness $150\,\mathrm{\mu}$m) with two freeform surfaces redirects the
  incoming turbulent wavefront onto an array of 91 single-mode fibers placed
  $50\,\mathrm{\mu}$m after the second surface.}
\label{fig:schematic}
\end{figure}

We consider a compact two-plane refractive system depicted in Fig.~\ref{fig:schematic},
consisting of a single glass slab of refractive index $n = 1.5$ and thickness
$150\,\mathrm{\mu}$m, whose two end facets are assumed to be freeform surfaces. The input
wavefront, corresponding to the image of the collecting pupil of diameter
$160\,\mathrm{\mu}$m, is placed $50\,\mathrm{\mu}$m before the first freeform surface. In
practice, this requires a telescope to image the physical collecting pupil, typically
several tens of centimeters to over one meter in diameter, onto the $160\,\mathrm{\mu}$m
input aperture of the device. The output single-mode fiber bundle is placed
$50\,\mathrm{\mu}$m after the second freeform surface, and its geometry corresponds to the
91-mode wide packing with concentric rings described in the previous section. The working
wavelength is $\lambda = 1.55\,\mathrm{\mu}$m.

\subsection{Optimized surfaces and physical interpretation}

The freeform surfaces are discretized on a $300 \times 300$ grid with sampling
$\Delta x = 1\,\mathrm{\mu}$m, as defined in the previous section, and optimized
by stochastic gradient descent on the loss function
\begin{equation}
  \mathcal{L} = 1 - \mathbb{E}_j\left[\frac{1}{M}\sum_{i=1}^{M}
  |\langle u_i | \psi_j \rangle|^2\right],
  \label{eq:loss}
\end{equation}
where $\{u_i\}_{i=1}^M$ are the $M = 91$ fiber bundle modes and $\psi_j$ is a
turbulent wavefront drawn from the ensemble of $N = 3000$ realizations generated
in the previous section at $D/r_0 = 8$. End-to-end optimization of the surface
profiles is performed using FluxOptics.jl~\cite{Barré:2026}, an open-source Julia
package for differentiable wave optics. At each iteration, the expectation in
Eq.~\eqref{eq:loss} is estimated over a batch of $N_b = 300$ randomly drawn
realizations from the ensemble:
\begin{equation}
  \hat{\mathcal{L}} = 1 - \frac{1}{M N_b} \sum_{i=1}^{M} \sum_{j=1}^{N_b}
  |\langle u_i | \psi_j \rangle|^2.
  \label{eq:loss_estimator}
\end{equation}
This loss function extends the multimode power coupling metric
introduced in~\cite{Barré:2022} to the case where the input fields
$\{\psi_j\}$ form a statistical ensemble of turbulent wavefront realizations
rather than an orthonormal basis, and where $N_b \neq M$.

\begin{figure}[htbp]
  \centering
  \includegraphics{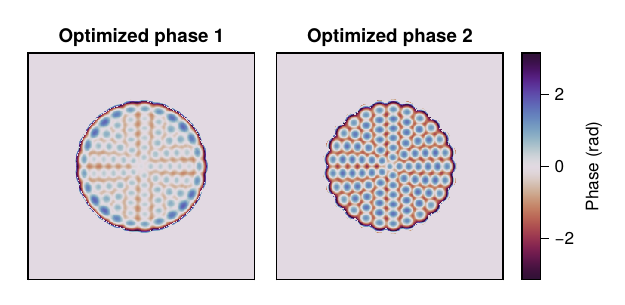}
  \caption{Optimized phase patterns of the two freeform surfaces of the glass slab.  The
    smooth, slowly-varying structure emerges spontaneously from the optimization on random
    turbulent inputs, reflecting the geometry of the output fiber bundle. Both panels span
    $300\,\mu\mathrm{m} \times 300\,\mu\mathrm{m}$.}
  \label{fig:optimized_phases}
\end{figure}

Optimization runs for 500 iterations, each processing a new batch, and converges
in 22 seconds on a GPU GeForce RTX 4070 Super. The resulting phase patterns,
shown in Fig.~\ref{fig:optimized_phases}, are remarkably smooth and slowly varying,
a non-trivial outcome given that the optimization was driven exclusively by random
turbulent inputs with no regularity constraint imposed on the surfaces. This
smoothness reflects the geometry of the output fiber bundle: the optimization
naturally produces a phase profile that locally redirects the incoming wavefront
toward each fiber core, in a manner reminiscent of a custom microlens arrangement.

\begin{figure}[htbp]
  \centering
  \includegraphics{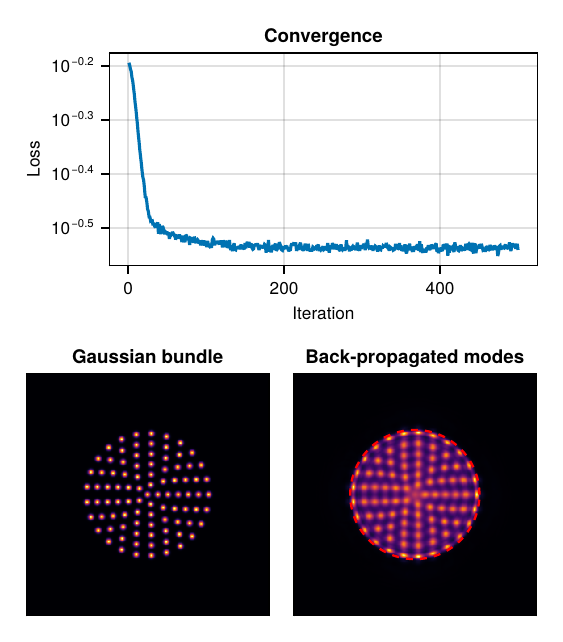}
  \caption{Top: convergence of the loss function over 500 SGD iterations.
    Bottom left: total intensity of the Gaussian fiber bundle in the output plane.
    Bottom right: total intensity of the Gaussian bundle modes back-propagated through
    the optimized device to the input pupil plane. The dashed red circle indicates
    the boundary the collecting pupil of diameter $D = 160\,\mathrm{\mu}$m.}
  \label{fig:optimization}
\end{figure}

To gain further insight into the optimized system, we back-propagate each Gaussian mode of
the fiber bundle through the device and examine the resulting field distribution at the
input pupil plane. As shown in Fig.~\ref{fig:optimization}, the back-propagated modes form a
dense arrangement of deformed Gaussian spots that collectively tile the collecting pupil,
strikingly similar to the dense packing basis of the previous section. This spontaneous
convergence toward a dense spatial coverage confirms the spatial support maximization
principle identified earlier: the optimization has redistributed the input sensitivity of
each fiber to cover the pupil as efficiently as possible, without any explicit instruction
to do so.

Table~\ref{tab:2_plane_system} summarizes the coupling performance of the optimized
system at $D/r_0 = 8$. With a mean coupling efficiency of $\bar{\eta} = 0.713$
without tip-tilt correction and $\bar{\eta} = 0.764$ with correction, the two-plane
system lies between the wide packing baseline and the LG and dense packing bases
reported in Table~\ref{tab:basis_comparison}. This result is consistent with the
output Gaussian bundle geometry, which corresponds to the wide packing layout and
therefore starts from a less favorable spatial coverage than the dense packing.
Tip-tilt correction reduces the coupling standard deviation from $\sigma = 0.053$
to $\sigma = 0.030$, following the same trend observed for the modal bases in the
previous section.

\begin{table}[htbp]
  \centering
  \caption{Mean power coupling efficiency $\bar{\eta}$ and standard deviation
  $\sigma$ for the optimized two-plane system, without and with tip-tilt
  correction, at $D/r_0 = 8$. This data extends  Table~\ref{tab:basis_comparison}.}
  \label{tab:2_plane_system}
  \begin{tabular}{lcccc}
    \toprule
    & \multicolumn{2}{c}{No correction}
    & \multicolumn{2}{c}{Tip-tilt corrected} \\
    \cmidrule(lr){2-3} \cmidrule(lr){4-5}
    & $\bar{\eta}$ & $\sigma$ & $\bar{\eta}$ & $\sigma$ \\
    \midrule
    2-plane system & 0.713 & 0.053 & 0.764 & 0.030 \\
    \bottomrule
  \end{tabular}
\end{table}

\subsection{Robustness of the optimized design across turbulence regimes}

\begin{figure}[htbp]
  \centering
  \includegraphics{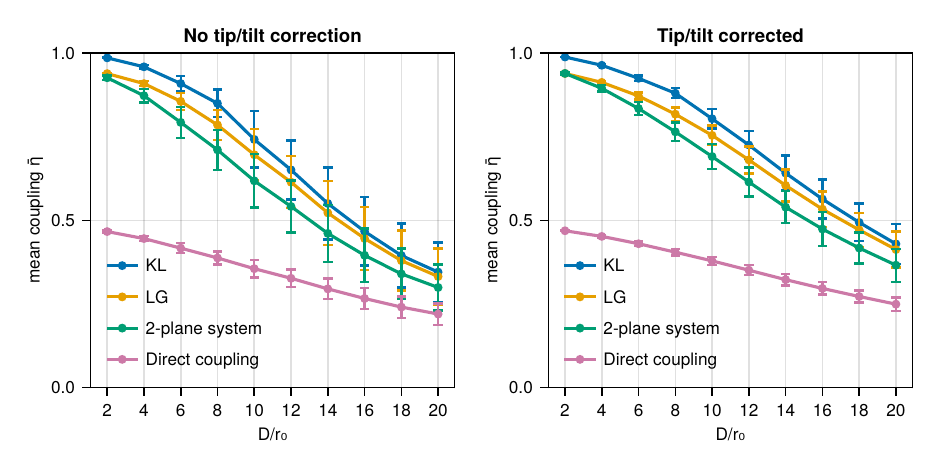}
  \caption{Mean power coupling efficiency $\bar{\eta}$ as a function of
  turbulence strength $D/r_0$ for the KL basis, LG basis,
  optimized two-plane system, and direct fiber coupling, without (left)
  and with (right) tip-tilt correction. The two-plane system is optimized
  at $D/r_0 = 8$. Error bars indicate one standard deviation over the
  turbulence ensemble.}
  \label{fig:robustness}
\end{figure}

The system was optimized at $D/r_0 = 8$ using a fixed ensemble of $N = 3000$ turbulent
realizations. Figure~\ref{fig:robustness} shows the mean coupling efficiency and its
standard deviation as a function of $D/r_0 \in [2, 20]$, compared to the KL upper bound, the
LG basis projection, and direct fiber coupling, both without and with tip-tilt correction.

The two-plane system remains competitive across the full range of turbulence conditions
considered. Without tip-tilt correction, the gap between the two-plane system and the KL
upper bound increases from $6\%$ at $D/r_0 = 2$ to approximately $17\%$ around
$D/r_0 = 10$--$12$, before slightly decreasing at higher turbulence strengths as all bases
become limited by the finite number of collected modes. The gap with respect to the ideal LG
projection follows a similar trend, ranging from $1.4\%$ at $D/r_0 = 2$ to $12\%$ at
$D/r_0 = 12$.  Tip-tilt correction substantially improves the relative performance of the
two-plane system, particularly at low turbulence strengths. At $D/r_0 = 2$, the corrected
two-plane system matches the LG projection to within $0.1\%$, effectively closing the
gap. At the design point $D/r_0 = 8$, the gap with KL is reduced from $16.6\%$ to $13.2\%$,
and the gap with LG from $9.8\%$ to $6.6\%$. For strong turbulence ($D/r_0 > 12$), the
benefit of tip-tilt correction diminishes, as higher-order wavefront aberrations
increasingly dominate over beam wander and cannot be compensated by tip-tilt alone. Overall,
tip-tilt correction offers a favorable trade-off in the moderate turbulence regime, where it
brings the two-plane system close to the performance of an ideal modal projection at minimal
added hardware complexity.

A notable observation concerns the coupling variance. Direct fiber coupling, despite its low
mean efficiency, exhibits a significantly lower standard deviation than all optimized bases
across the entire turbulence range, particularly at high $D/r_0$. This is a manifestation of
the spatial diversity effect: non-overlapping fiber modes sample independent regions of the
pupil and average out turbulence fluctuations, at the cost of reduced mean power
collection. Conversely, the KL basis, while optimal in mean, exhibits the highest variance
under strong turbulence, as its modes are concentrated in the regions of highest average
energy and therefore most sensitive to turbulent fluctuations. The two-plane system occupies
an intermediate position, with a variance comparable to the LG basis.

These results suggest that mean coupling efficiency and coupling variance are competing
objectives under strong turbulence, and that the current design, optimized solely for mean
power, leaves room for improvement in fading statistics. A natural extension would be to
include a variance penalty term in the loss function~\cite{Barré:2018}, which could further
reduce fading without sacrificing mean efficiency.

It is worth noting that the two-plane architecture imposes no constraint on the fiber bundle
geometry: the optimizer freely adapts to any spatial arrangement of output modes, whether
concentric rings as used here, Cartesian grids, or arbitrary configurations dictated by
fabrication constraints. The concentric ring layout was chosen for its circular symmetry and
to match the number of modes of the Laguerre-Gaussian basis, but simulations with Cartesian
geometries and non-standard mode counts converge equally well, suggesting that the approach
is directly applicable to photonic lanterns or multicore fibers of arbitrary design. From a
practical standpoint, the two-plane geometry is already well suited for laboratory
implementation with a spatial light modulator, while fs-laser written photonic lanterns with
dense core packing represent a promising path toward compact, alignment-free hardware
embodiments of the same design principle.

\section{Conclusion}

We have presented a systematic comparison of modal bases for turbulent wavefront power
collection in free-space optical communications, followed by the inverse design of a compact
two-plane refractive demultiplexer optimized directly over atmospheric turbulence
statistics.

The modal basis comparison reveals that the spatial coverage of the collecting pupil is the
primary determinant of power collection efficiency, independently of the specific basis
chosen. A dense packing of Gaussian modes, after re-orthonormalization, performs as well as
an optimal Laguerre-Gaussian basis, and the functional space it spans is analogous to that
of fs-laser written photonic lanterns, where closely-spaced fiber cores produce coupled
modes that are not individually controlled yet collectively cover the input
aperture. Conversely, a sparse orthogonal arrangement of non-overlapping fiber modes
achieves only half the coupling efficiency of denser bases.

The optimized two-plane system, designed without any prescribed modal decomposition of the
turbulent wavefront, approaches the performance of an ideal Karhunen-Loève projection and
significantly improves over direct fiber coupling. Remarkably, the optimized phase patterns
are perfectly smooth and respect the geometry of the output fiber bundle, bearing a strong
resemblance to a custom microlens arrangement despite having been optimized exclusively on
random turbulent inputs. This demonstrates that constraining the demultiplexer to decompose
the incoming field onto a prescribed basis, as in MPLC-based approaches targeting a
Hermite-Gaussian decomposition, imposes an unnecessary design burden: a two-plane geometry
is sufficient to capture most of the available gain when the optimization criterion is
matched to the turbulence statistics and the spatial support of the collected modes covers
the pupil efficiently. Back-propagating the fiber bundle modes through the optimized device
shows that each fiber mode is mapped to a compact spot in the pupil plane, collectively
tiling the available aperture, confirming the spatial support maximization principle
identified in the basis comparison. Tip-tilt correction further narrows the performance gap,
particularly in the mild turbulence regime.

A key finding of the robustness analysis is that mean coupling efficiency and coupling
variance are competing objectives under strong turbulence. The current design, optimized
solely for mean power, exhibits higher fading than direct fiber coupling despite its
superior mean efficiency, pointing to a natural extension: incorporating a variance penalty
into the loss function~\cite{Barré:2018} to simultaneously improve mean efficiency and reduce
fading. From a practical perspective, the two-plane geometry is well suited for laboratory
validation with a spatial light modulator, and fs-laser written photonic lanterns with dense
core packing represent a particularly promising hardware embodiment of the spatial support
maximization principle demonstrated here. Coherent combination of the fiber outputs on a
photonic integrated circuit~\cite{demarinis2025} would further complete the receiver chain
and enable direct comparison with full system performance metrics.

\paragraph{Funding.}
This research received no external funding.

\paragraph{Data availability.}
All simulations were performed using FluxOptics.jl~\cite{Barré:2026}, an open-source Julia
package for differentiable wave optics. Data and code underlying the results presented in
this paper are available from the author upon reasonable request.

\paragraph{AI assistance.}
The author used an AI language model (Claude, Anthropic) as an assistant during the
preparation of this manuscript, for tasks including literature search, LaTeX drafting, and
technical discussion.  All scientific content, simulations, results, and conclusions were
developed independently by the author.

\let\OLDthebibliography\thebibliography
\renewcommand\thebibliography[1]{
  \OLDthebibliography{#1}
  \setlength{\parskip}{1ex}
  \setlength{\itemsep}{0pt plus 1ex}
}

\bibliographystyle{unsrturl}
\bibliography{FSO}

\end{document}